\documentclass[prd,twocolumn,showpacs,amsmath,amssymb,nofootinbib]{revtex4-2}
\usepackage{epsf}
\usepackage{graphicx}
\usepackage{epsfig}
\usepackage{xcolor}
\usepackage{subfigure}
\usepackage{pstricks}
\usepackage{pst-node}
\usepackage{rotating}
\usepackage{graphics}
\usepackage{latexsym}
\usepackage{dcolumn}
\usepackage{bm}
\usepackage{times}
\usepackage{indentfirst}
\usepackage{float}
\usepackage{overpic}
\usepackage{amsmath}
\usepackage{mathrsfs}
\usepackage{appendix}
\usepackage{upgreek}
\usepackage{chngcntr}
\usepackage{amssymb}%
\usepackage{pifont}%
\usepackage{threeparttable}
\usepackage[top=0.5in,bottom=0.8in,left=0.9in,right=0.9in]{geometry}
\usepackage{subfigure}
\usepackage{lineno}
\usepackage{multirow}
\usepackage{makecell}
\usepackage{lipsum}
\usepackage{hyperref}

\usepackage{besphysics}
\hypersetup{
	colorlinks=true,
	linkcolor=blue,
	filecolor=blue,
	urlcolor=blue,
	citecolor=blue,
}

\include{def-com}

\begin{document}
%\linenumbers	

\title{\boldmath{Measurement of $\ee\to\etajpsi$ Cross Section from $\sqrt{s}=3.808\gev$ to $4.951\gev$}}

%% Saved at => 2023-07-27
\author{
\begin{small}
\begin{center}
M.~Ablikim$^{1}$, M.~N.~Achasov$^{4,b}$, P.~Adlarson$^{75}$, X.~C.~Ai$^{81}$, R.~Aliberti$^{35}$, A.~Amoroso$^{74A,74C}$, M.~R.~An$^{39}$, Q.~An$^{71,58}$, Y.~Bai$^{57}$, O.~Bakina$^{36}$
% ...
\\
\vspace{0.2cm}
(BESIII Collaboration)\\
\vspace{0.2cm} {\it
$^{1}$ Institute of High Energy Physics, Beijing 100049, People's Republic of China\\
% ...
$^{4}$ Budker Institute of Nuclear Physics SB RAS (BINP), Novosibirsk 630090, Russia\\
% ...
$^{35}$ Johannes Gutenberg University of Mainz, Johann-Joachim-Becher-Weg 45, D-55099 Mainz, Germany\\
$^{36}$ Joint Institute for Nuclear Research, 141980 Dubna, Moscow region, Russia\\
% ...
$^{39}$ Liaoning Normal University, Dalian 116029, People's Republic of China\\
% ...
$^{57}$ Southeast University, Nanjing 211100, People's Republic of China\\
$^{58}$ State Key Laboratory of Particle Detection and Electronics, Beijing 100049, Hefei 230026, People's Republic of China\\
% ...
$^{71}$ University of Science and Technology of China, Hefei 230026, People's Republic of China\\
% ...
$^{74}$ University of Turin and INFN, (A)University of Turin, I-10125, Turin, Italy; (B)University of Eastern Piedmont, I-15121, Alessandria, Italy; (C)INFN, I-10125, Turin, Italy\\
$^{75}$ Uppsala University, Box 516, SE-75120 Uppsala, Sweden\\
% ...
$^{81}$ Zhengzhou University, Zhengzhou 450001, People's Republic of China\\
\vspace{0.2cm}
% ...
$^{b}$ Also at the Novosibirsk State University, Novosibirsk, 630090, Russia\\
% ...
}
\end{center}
\vspace{0.4cm}
\vspace{0.4cm}
\end{small}
}
%% ends here %%
	
\date{\today}

\renewcommand{\abstractname}{}
\begin{abstract}
Using data samples with an integrated luminosity of 22.42\,$\ifb$
collected by the BESIII detector operating at the BEPCII storage ring,
we measure the cross sections of the $\ee\to\etajpsi$ process at
center-of-mass energies from 3.808 to 4.951~\gev.  Three structures
are observed in the line shape of the measured cross sections.  A
maximum-likelihood fit with $\psi(4040)$, two additional resonances,
and a non-resonant component is performed.  The mass and width of
the first additional state are $(4219.7\pm2.5\pm4.5)\mevcc$ and
$(80.7\pm4.4\pm1.4)\mev$, respectively, consistent with the
$\psi(4230)$.  For the second state, the mass and width are
$(4386\pm13\pm17)\mevcc$ and $(177\pm32\pm13)\mev$, respectively,
consistent with the $\psi(4360)$.  The first uncertainties are
statistical and the second ones are systematic.  The statistical
significance of $\psi(4040)$ is $8.0\sigma$ and those for $\psi(4230)$
and $\psi(4360)$ are more than $10.0\sigma$. 
\end{abstract}

\maketitle


\section{INTRODUCTION} \label{}

Hadron spectroscopy is a fascinating field full of discoveries and
surprises.  Over the past decades, many charmonium-like states
with $J^{PC} = 1^{--}$, called $Y$ states, have been discovered and
confirmed by numerous experiments.  As non-standard hadron candidates
beyond the conventional quark model, these states have many
characteristics that are different from the traditional ones and have
stimulated great interests both experimental and theoretical.  The
masses of these $Y$ states are above $D\bar{D}$ threshold, and they have
strong coupling to hidden-charm final states.  Many theoretical
interpretations, such as hybrid mesons, compact tetraquark states and
hadronic molecules~\cite{BRAMBILLA20201}, have been proposed.
However, none of them can account for all unusual properties of these $Y$
states. 

Among these exotic states, $\psi(4230)$, previously known as
$Y(4260)$, and $\psi(4360)$, previously known as $Y(4360)$, were first
discovered by BaBar and Belle using initial-state-radiation (ISR) in
the $\ee\to\gamma_{\rm
ISR}\pppm\jpsi$~\cite{PRL95142001,PRL99182004,PRD86051102,PRL110252002}
and $\ee\to\gamma_{\rm
  ISR}\pppm\psi(3686)$~\cite{PRL98212001,PRD89111103,PRL99142002,PRD91112007}
  processes.  With higher statistics achieved by BESIII, they are
  observed via more processes and measured with improved precision. The
  $\psi(4230)$ is observed in the
  $\ee\to\pppm\jpsi$\cite{PRL118092001,BESIII:2022qal},
  $\ee\to\piz\piz\jpsi$~\cite{BESIII:2020oph}, $\ee\to
  K^0_SK^0_S\jpsi$~\cite{PhysRevD.107.092005}, $\ee\to K^+ K^-
  \jpsi$~\cite{BESIII:2022joj}, $\ee\to
  \pppm\psi(3686)$~\cite{PRD96032004,PRD104052012}, $\ee\to\pppm
  h_c$\cite{PRL118092002}, $\ee\to\omega
  \chi_{c0}$~\cite{PRL114092003,PRD99091103}, and
  $\ee\to\pi^+D^0D^{*-}$~\cite{BESIII:2018iea} processes, and the
  $\psi(4360)$ is observed in the
  $\ee\to\pppm\psi(3686)$~\cite{PRD96032004,PRD104052012},
  $\ee\to\pppm h_c$~\cite{PRL118092002}, and
  $\ee\to\pppm\psi_2(3823)$~\cite{PhysRevLett.129.102003} processes.
  The parameters of each of these two resonances, such as their masses and
  widths, are similar, but there are still differences between these
  decay modes. 

In recent years, the branching fractions, partial decay widths, and
the quark components of $\psi(4230)$ and $\psi(4360)$ have been
predicted by many theoretical models.
Assuming the $\psi(4230)$ is a conventional $\psi(4S)$ state and using
the electronic partial widths provided by
Refs.~\cite{Dong:1994zj,Li:2009zu}, the upper limit of the
$\psi(4S)\to\eta\jpsi$ branching fraction is predicted to be
$1.9\times 10^{-3}$~\cite{PRD91094023}. Assuming that $\psi(4360)$ is
a pure $4(3^3D_1)$ state, the partial width of the
$\psi(4360)\to\eta\jpsi$ decay is estimated in
Ref.~\cite{PRD95114031}. Assuming a hadronic molecular state,
Refs.~\cite{PRD88094008,PhysRevD.96.054017} predict the partial
decay widths or branching fractions of $\eta\jpsi$ and other final
states for $\psi(4230)$ and $\psi(4360)$ decays.

In this paper, an updated analysis of $\ee\to\etajpsi$ at 44
center-of-mass (c.m.) energies between $3.808$ and $4.951$ GeV is
performed, using a similar approach as in Ref.~\cite{PRD102031101}.
Additional data samples on the side of the $\psi(4230)$ peak and from
$4.612$ to $4.951\gev$ are used, which can describe $\psi(4230)$,
$\psi(4360)$ and the non-resonance more precisely than the previous
measurement~\cite{PRD102031101}, as well as allow a search for heavier
$Y$ states in the $\ee\to\eta\jpsi$ process. In this
analysis, the $\jpsi$ is reconstructed via
$\jpsi\to\LL (\ell=e/\mu)$ and the $\eta$ is reconstructed via
$\eta\to\GG$ (Mode I) and $\eta\to\piz\pppm$ (Mode II).

\section{THE BESIII DETECTOR AND DATA SAMPLES} 

The BESIII detector~\cite{NIMA614345399} records symmetric $e^+e^-$
collisions provided by the BEPCII storage ring~\cite{Yu:2016cof}.  The
cylindrical core of the BESIII detector covers $93\%$ of the full
solid angle and consists of a helium-based multilayer drift
chamber~(MDC), a plastic scintillator time-of-flight system~(TOF), and
a CsI(Tl) electromagnetic calorimeter~(EMC), which are all enclosed in
a superconducting solenoidal magnet providing a 1.0~T magnetic
field. The solenoid is supported by an octagonal flux-return yoke with
resistive plate counter muon identification modules interleaved with
steel.  The charged-particle momentum resolution at $1~{\rm GeV}/c$ is
$0.5\%$, and the specific ionization energy loss ${\rm d}E/{\rm d}x$
resolution is $6\%$ for electrons from Bhabha scattering. The EMC
measures photon energies with a resolution of $2.5\%$ ($5\%$) at
$1\gev$ in the barrel (end cap) region. The time resolution in the TOF
barrel region is 68~ps, while that in the end cap region was
110~ps. The end cap TOF system was upgraded in 2015 using multi-gap
resistive plate chamber technology, providing a time resolution of
60~ps~\cite{Cao:2020ibk}.

The data samples used in this analysis are listed in
Appendix~\ref{app:cs}. Among them, 76.05\% are
collected after the end cap TOF upgrade.  The c.m. energy is measured
using dimuon events with a precision of $0.8\mev$ for data samples
with $\sqrt{s}$ lower than
$4.612\gev$~\cite{BESIII:2015zbz,CPC45103001} and using
$\Lambda_{c}^{+}\bar{\Lambda}_{c}^{-}$ events with a precision of
$0.6\mev$ for data samples with $\sqrt{s}$ higher than or equal to
$4.612\gev$~\cite{luminosity2ecms}. The integrated luminosity is
determined with an uncertainty of $1.0\%$ by analyzing large-angle
Bhabha scattering
events~\cite{luminosity0,luminosity1,luminosity2ecms}.

Monte Carlo (MC) simulation is used to optimize event selection
criteria, estimate background, and determine event selection
efficiencies.  The BESIII MC simulation framework is based on {\sc
geant4}~\cite{NIMA506250303} and includes the geometric
description~\cite{Huang:2022wuo} of the BESIII detector and the
realistic representation of the electronic readout.  The beam energy
spread and ISR in the $e^+e^-$ annihilation are modeled with the {\sc
kkmc} generator~\cite{Jadach:1999vf,Jadach:2000ir}.  The inclusive MC
sample includes the production of open charm processes, the ISR
production of vector charmonium(-like) states, and the non-resonant
processes incorporated in {\sc kkmc}.  All particle decays are modeled
with {\sc evtgen}~\cite{Lange:2001uf,Ping:2008zz} using branching
fractions either taken from the Particle Data Group
(PDG)~\cite{Workman:2022ynf}, when available, or otherwise estimated
with {\sc lundcharm}~\cite{Chen:2000tv,Yang:2014vra}, {\sc
ConExc}~\cite{CPC38083001} and {\sc
phokhara}~\cite{Campanario:2019mjh}.  Final state radiation from
charged final state particles is incorporated using {\sc
photos}~\cite{Richter-Was:1992hxq}.  Signal MC samples of
$\ee\to\etajpsi$ with the corresponding $\jpsi$ and $\eta$ decay modes
are generated using {\sc helamp}~\cite{Lange:2001uf} with parameters
(1 0 0 0 -1 0) and {\sc evtgen} at each c.m. energy.  ISR is simulated
with {\sc kkmc}, and the maximum energy of the ISR photon is adjusted
according to the $\etajpsi$ mass threshold.

\section{EVENT SELECTION \label{eventsel}} 

The good charged tracks are required to be within the angle coverage
of the MDC, $|\!\cos\theta|<0.93$, where the $\theta$ is defined with
respect to the $z$ axis, which is the symmetry axis of the MDC. The
distance of closest approach to the $\ee$ interaction point (IP) must
be less than 1~cm in the transverse plane, $|V_{xy}|<1$~cm, and less
than 10~cm along the $z$ axis, $|V_z|<10$~cm.  Photon candidates are
identified using showers in the EMC.  The deposited energy of each
shower must be more than $25\mev$ in the barrel region
($|\!\cos\theta|<0.80$) and more than $50\mev$ in the end cap region
($0.86<|\!\cos\theta|<0.92$).  To suppress electronic noise and
showers unrelated to the event, the difference between the EMC time
and the event start time is required to be within $[0, 700]$ ns. To
remove photons produced by interactions of charged tracks, the opening
angle between a shower and its nearest charged track has to be greater
than $20^\circ$.  Candidate events are required to have two (Mode I)
or four (Mode II) charged tracks with zero net charge and at least two
photons.

For signal candidates, the pions and leptons are distinguished by
their momenta. The charged tracks with momenta above $1.0\gevc$ are
assigned to be leptons, while others are assumed to be pions.
The separation of electrons and muons is accomplished using the
deposited energy ($E$) in the EMC.  Muons must satisfy $E\leq0.4\gev$,
while electrons must satisfy $E/pc>0.8$, where $p$ is the momentum of
the charged
track. For Mode I, signal candidate events are required to have a
lepton pair with the same flavor and opposite charge, and at least two
photons. For Mode II, two additional pions with opposite charge are
required.
	
To improve the resolution and suppress the background for Mode I, a
four-constraint (4C) kinematic fit imposing energy-momentum
conservation is performed under the hypothesis
$\ee\to\gamma\gamma\LL$.  For Mode II, a five-constraint (5C)
kinematic fit is performed under the hypothesis
$\ee\to\gamma\gamma\pppm\LL$ with an additional $\piz$ mass constraint
for the photon pair. For candidate events with more than two photons, 
the combination with the smallest $\chi^2_{\rm 4C}$ or
$\chi^2_{\rm 5C}$ of the 4C or 5C kinematic fit is retained. We
require $\chi^2_{\rm 4C}<40$ for Mode I and $\chi^2_{\rm 5C}<80$ for
Mode II.  For Mode I, to suppress the background from radiative Bhabha
and dimuon processes, the energy of each selected photon after the 4C
kinematic fit is required to be greater than 0.08$\gev$.

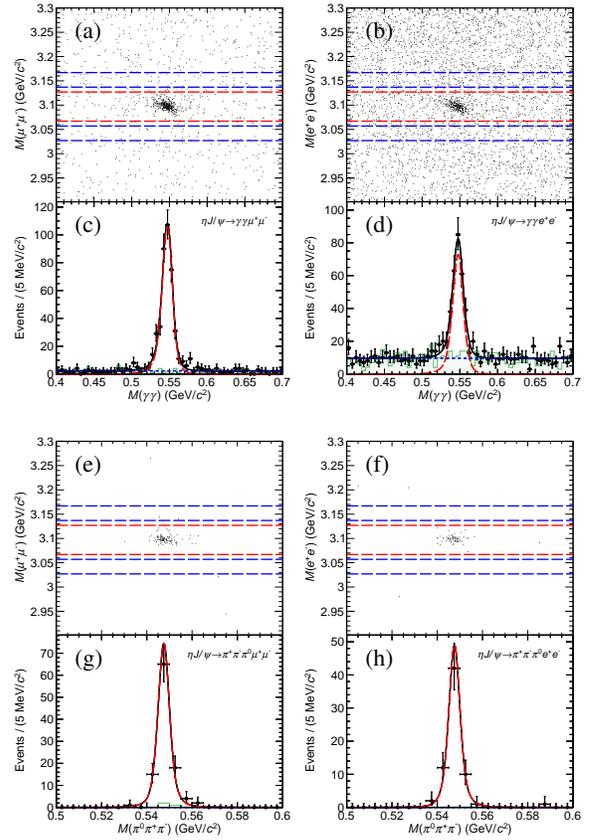
\begin{figure}[!htp]
    \begin{center}
           \begin{overpic}[width=1.5in]{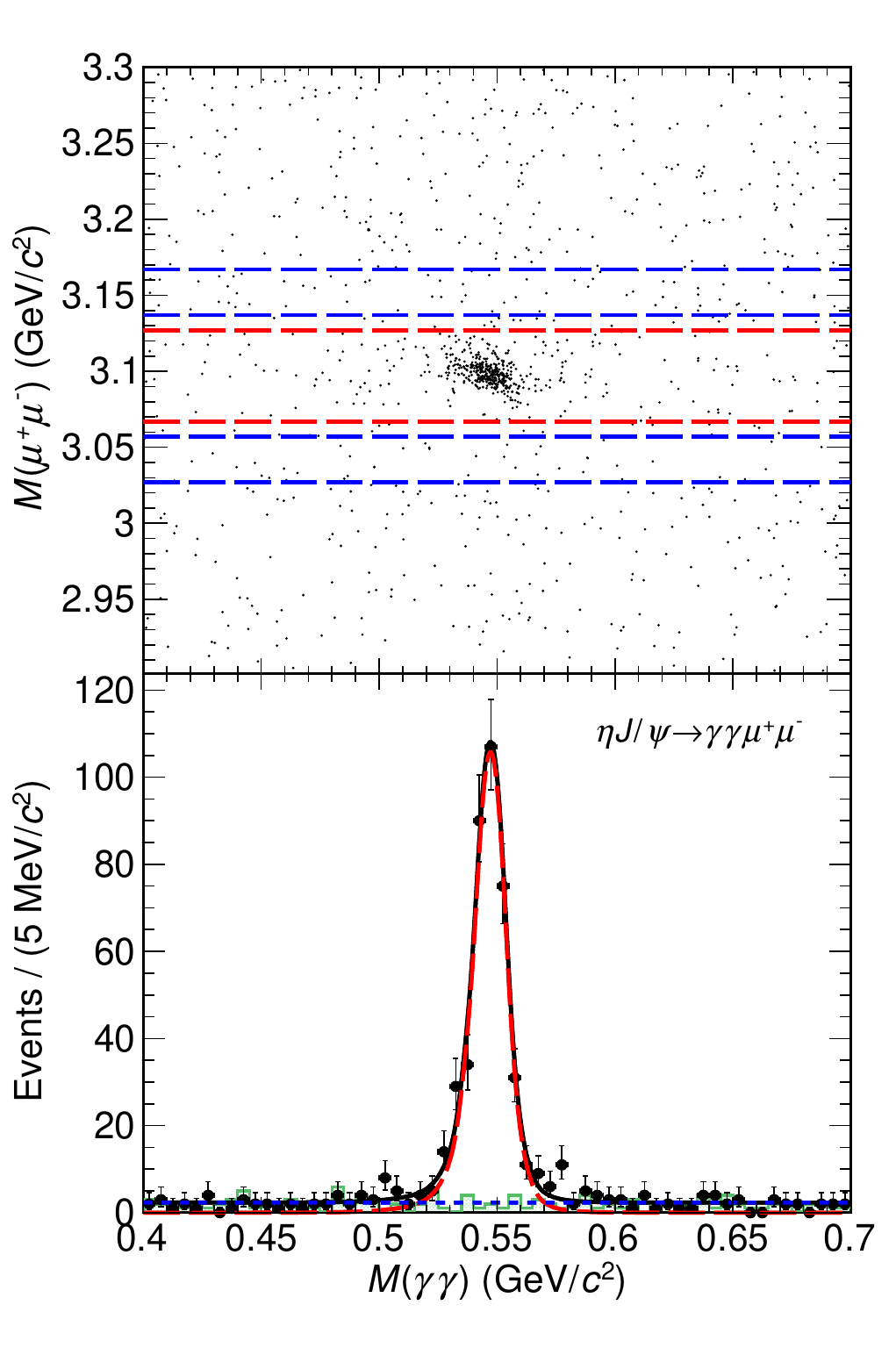}
           \put(15,88){(a)}
           \put(15,43){(c)}
           \end{overpic}   
           \begin{overpic}[width=1.5in]{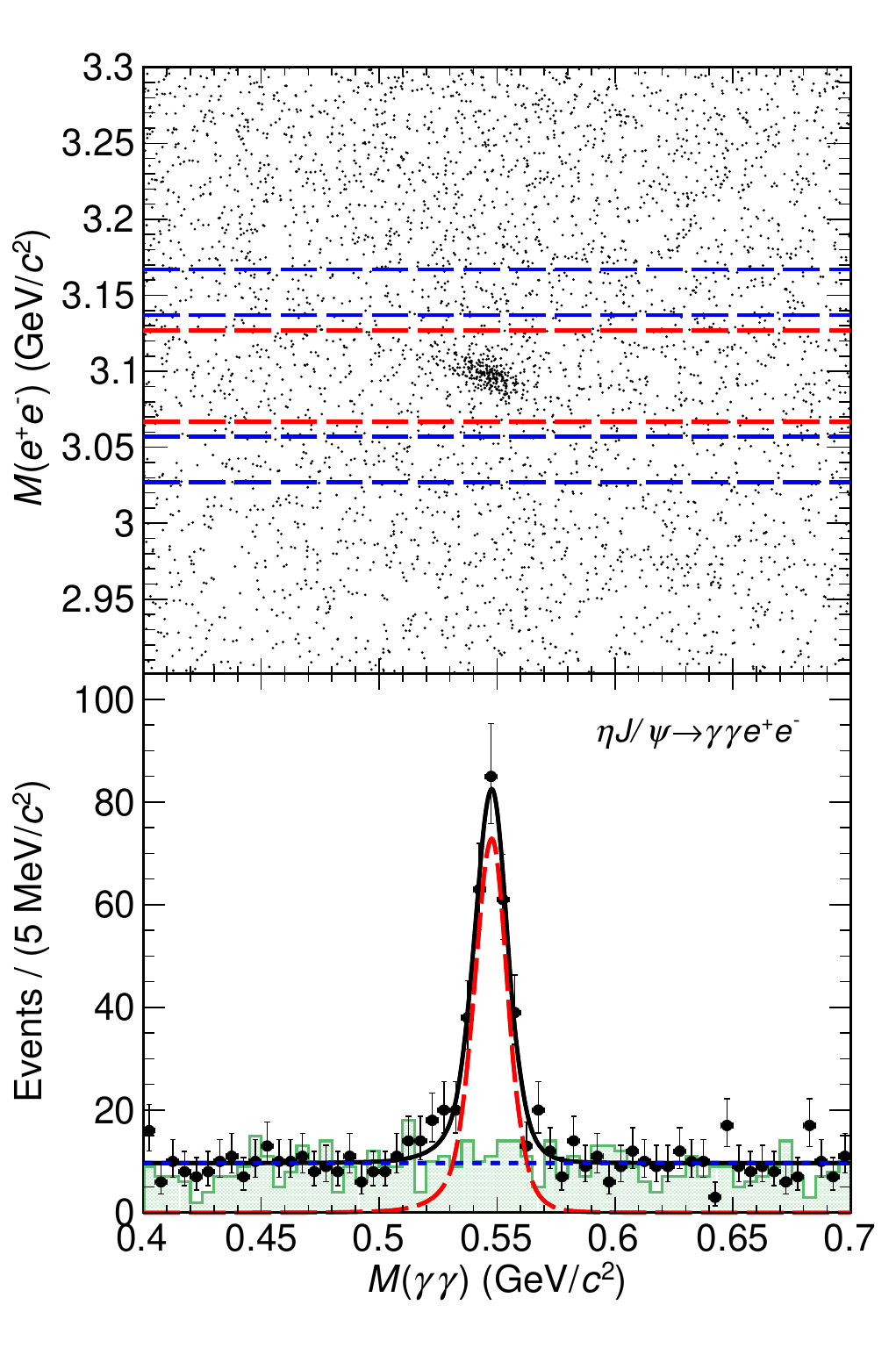} 
           \put(15,88){(b)}
           \put(15,43){(d)}
           \end{overpic}   
           \begin{overpic}[width=1.5in]{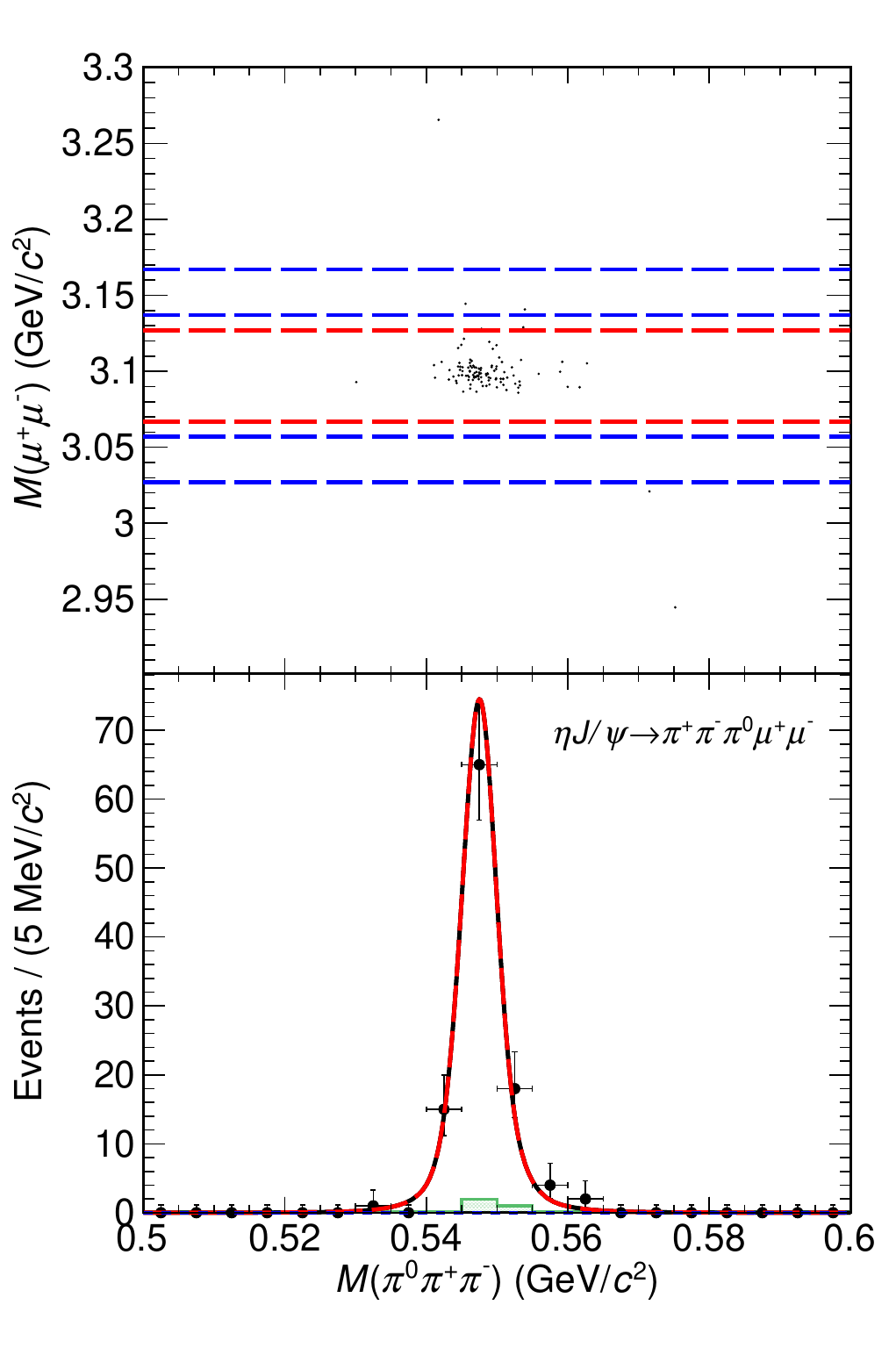} 
           \put(15,88){(e)}
           \put(15,43){(g)}
           \end{overpic}   
           \begin{overpic}[width=1.5in]{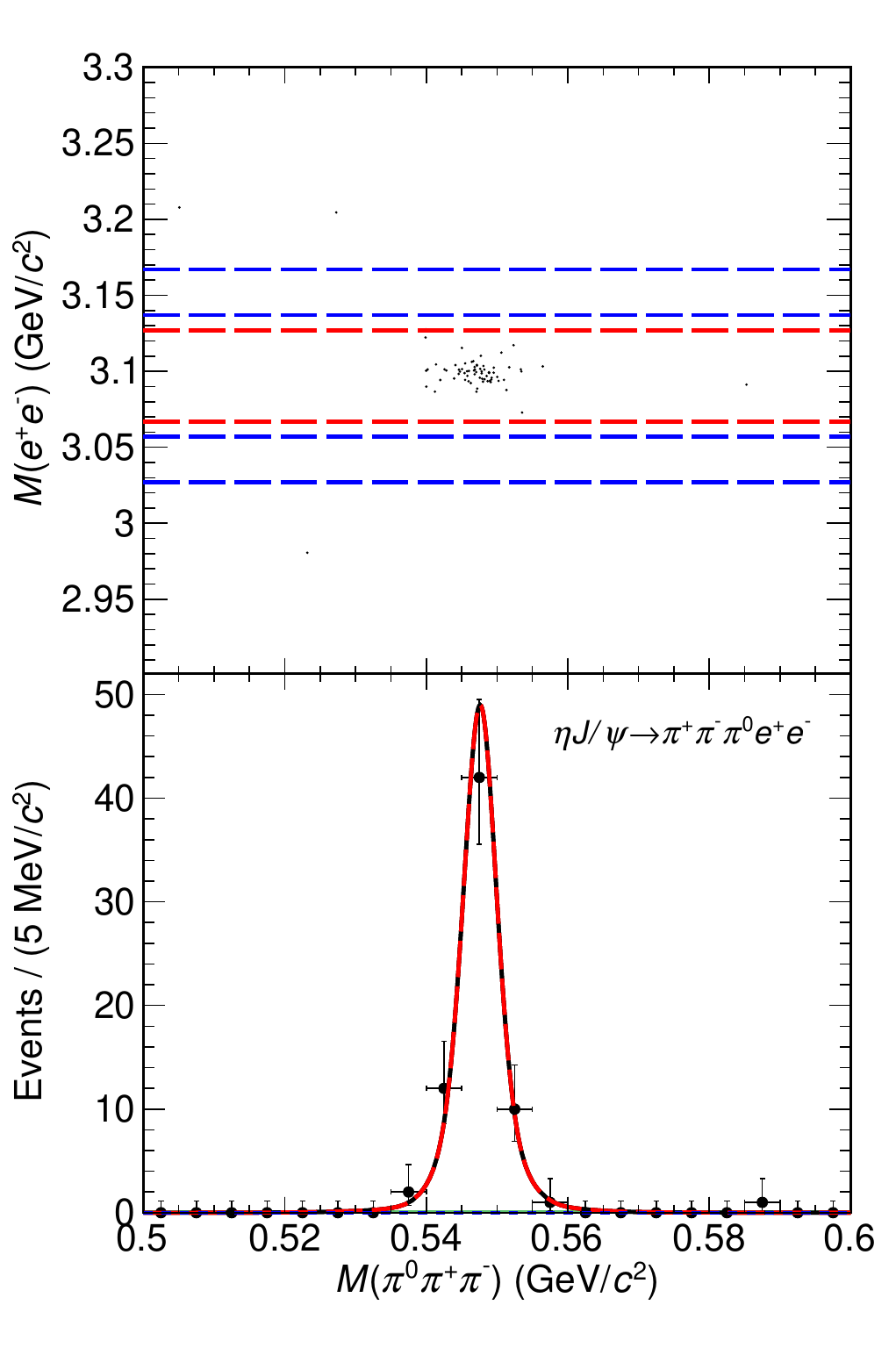} 
           \put(15,88){(f)}
           \put(15,43){(h)}
           \end{overpic}
   \end{center} \caption{(a), (b), (e), and (f) are the distributions
   of $M(\LL)$ versus $M(\gamma\gamma / \piz\pppm)$, where the region
   enclosed by red~(blue) long-dashed lines is the signal~(sideband)
   region. (c), (d), (g), and (h) are the distributions of
   $M(\gamma\gamma/\piz\pppm)$ in the $\jpsi$ signal region of the data
   sample at $\sqrt s = 4.226\gev$, where the dots with error bars
   show data, the green histogram shows the events from the $J/\psi$ mass
   sideband, the black solid, red long-dashed and blue short-dashed
   lines denote the fit result, signal, and background,
   respectively. The top~(bottom) four panels correspond to Mode I~(Mode
   II).}

        \label{fig:dots_4fit}
\end{figure}

Figure~\ref{fig:dots_4fit} shows the distributions of the invariant
mass of the $\LL$ pair ($M(\LL)$) versus those of the $\gamma\gamma$
pair ($M(\gamma\gamma)$) or $\piz\pppm$ ($M(\piz\pppm)$) for selected
events of the data sample at $\sqrt{s} = 4.226\gev$. A clear enhancement
from the signal events appears at the intersection of the $\jpsi$ and
$\eta$ mass regions in data. Because of the much larger cross section
of the radiative Bhabha process, the background in $\jpsi\to\ee$ is more
serious than for $\jpsi\to\mu^+\mu^-$ in Mode I. Signal candidates are
required to be within the $\jpsi$ mass region, defined as $[3.067,
3.127]\gevcc$ on $M(\ell^+\ell^-)$.  The events in the $\jpsi$ mass
sideband regions, defined as $[3.027, 3.057]$ and $[3.137,
3.167]\gevcc$, are used to estimate the non-$\jpsi$ background, and no
peaking background is observed in the $M(\gamma\gamma)$ or
$M(\piz\pppm)$ distributions.

\section{CROSS SECTION MEASUREMENT}

The Born cross section $\sigma^{\rm{B}}$ is determined by
\begin{equation} \label{eq:born_xsec} \begin{aligned} \sigma^{\rm B} =
\frac{N_{\rm obs}}{\mathcal{L}_{\rm
        int}\cdot(1+\delta^{\rm
        ISR})\cdot\frac{1}{|1-\Pi|^2}\cdot\mathcal{B}\cdot\epsilon},
        \end{aligned} \end{equation} 

\noindent where $N_{\rm obs}$ is the signal yield, $\mathcal{L}_{\rm
int}$ is the integrated luminosity, $(1+\delta^{\rm ISR})$ is the ISR
correction factor,
        $\frac{1}{|1-\Pi|^2}$ is the vacuum polarization factor taken
       from Ref.~\cite{WorkGroupRadiativeCorrections:2010bjp},
        $\mathcal{B}$ is the product of the branching fractions of the
        intermediate states in the subsequent decays from the
        PDG~\cite{Workman:2022ynf}, and $\epsilon$ is the signal
        detection efficiency. The
        ISR correction factor and the detection efficiency are
        estimated based on signal MC samples, and weighted by a
        dressed cross section iterative weighting
        method~\cite{FPB1664501}. The relationship between the dressed
        cross sections and the Born cross sections is described by
        $\sigma^{\rm{dressed}}= \frac{\sigma^{\rm B}}{|1-\Pi|^2}$.

\begin{figure*}
   \includegraphics[width=4.40in]{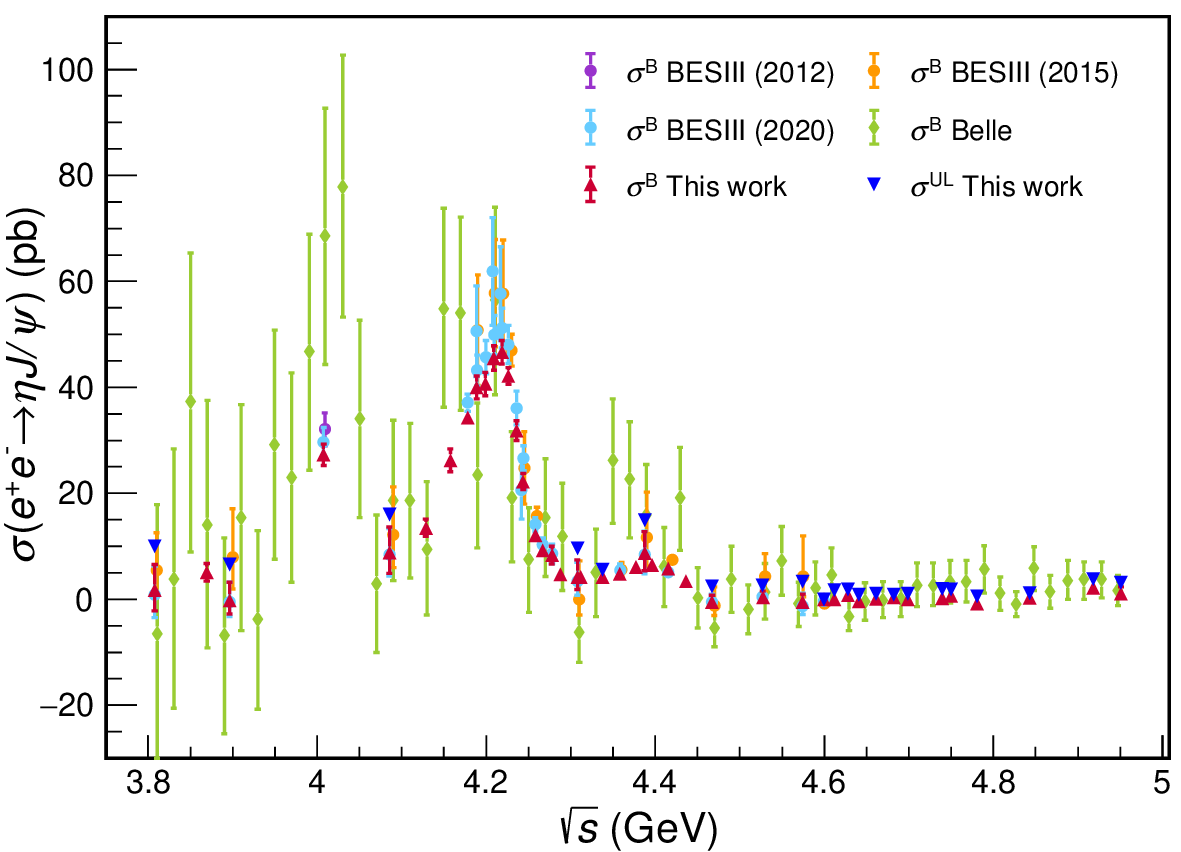}
   \caption{The Born cross sections and upper limits of
   $\ee\to\eta\jpsi$. The purple, orange and pale blue dots with error
   bars are the Born cross sections measured previously at
   BESIII~\cite{PRD86071101,PRD91112005,PRD102031101}; the green
   diamonds with error bars are the Belle results~\cite{PRD87051101};
   and the red triangles with error bars and the blue inverted triangles
   are the nominal Born cross sections and the upper limits from this
   work ($\sigma^{UL}$). The errors shown are the quadratic sum of
   statistical and systematic uncertainties.\label{fig:ls_cmp}}
   \end{figure*}

The measured cross section for each c.m. energy is obtained by a
simultaneous unbinned maximum-likelihood fit to the $M(\gamma\gamma)$
and $M(\piz\pppm)$ spectra 
extracted from the $J/\psi\to e^+e^-$ and $J/\psi\to\mu^+\mu^-$ 
modes separately, 
where the cross section is the shared
parameter between the four studied final states. The signal shape is
described by a simulated shape convolved with a Gaussian function,
which accounts for the difference of resolution between data
and MC simulation. Among different data samples, the parameters of
this Gaussian function are common and 
fixed to two different sets of values for Mode I and Mode II.  To
determine the parameters, simultaneous fits to the $M(\gamma\gamma)$
and $M(\piz\pppm)$ are performed, using data samples
with large statistics ($\sqrt{s} = 4.178, 4.209, 4.219, 4.226,
4.258~\mathrm{and}~4.416\gev$).  The background shape is described by
a linear function.  For those collision energy points where the
statistics are insufficient to observe a significant signal, an upper
limit for the cross section at the 90\% confidence level is
determined, taking into account the systematic
uncertainties~\cite{Cranmer:2000du,Liu_2015,stenson2006exact}.  The
measured cross sections are shown in Fig.~\ref{fig:ls_cmp}.  Our
measurements are in good agreement with earlier results from
BESIII~\cite{PRD86071101,PRD91112005,PRD102031101} and
with Belle~\cite{PRD87051101}.  The small differences
between this analysis and the previous BESIII results are due to 
increased statistics of data samples with the same c.m. energy and the updated
parameters of the {\sc helamp} generator. The cross sections
and quantities used for their measurements are summarized in
Appendix~\ref{app:cs}.

\section{FIT TO THE CROSS SECTION}

To search for possible resonances in the $\ee\to\eta\jpsi$ process, a
maximum-likelihood fit is performed to the dressed cross sections. For
data samples with large statistics, the likelihood is constructed by
assuming that the cross section obeys a normalized Gaussian
distribution, denoted as $G_i$, whose mean and standard deviation are estimated
by the measured cross section and the corresponding statistical
uncertainties.  But in cases where the data sample
lacks a significant signal, the likelihood is constructed through the
utilization of the Bayesian method for upper limit calculation, denoted as $L_{j}^{\rm scan}$.
The numbers of the data samples with large statistics or lacking a
significant signal are denoted as $N_1$ and $N_2$, respectively.
Therefore, the likelihood function of all data samples is composed of
two parts as

\begin{equation} \label{eq:llh_func_nominal}
\begin{aligned}
        L = \prod\limits_{i=1}^{N_1}G_{i} \cdot \prod\limits_{j=1}^{N_2}L_{j}^{\rm scan}.
\end{aligned}
\end{equation}

The fit function is parameterized as a coherent sum of three
Breit-Wigner functions, describing the structures around 4040, 4220 and
4390$\mevcc$, and a non-resonant component: 
\begin{equation} \label{eq:ls_func} \begin{aligned}
        \sigma^{\rm dressed}_{\rm fit}(\sqrt{s}) = |\sqrt{\sigma_{NY}(\sqrt{s})} + BW_1(\sqrt{s})e^{i\phi_1} \\
        + BW_{2}(\sqrt{s})e^{i\phi_2} + BW_3(\sqrt{s})e^{i\phi_3}|^2,
        \end{aligned} \end{equation} 

\noindent where the $\phi_i$ are the relative phases between three
resonances and the non-resonant, $BW_i$ is the Breit-Wigner function with the two-body phase space factor $\Phi(\sqrt{s})$: 

\begin{equation}
        \label{eq:bw_func} \begin{aligned}
        BW_i(\sqrt{s}) = \frac{\sqrt{12\pi\mathcal{B}_i\Gamma^{\ee}_i\Gamma_i}}{s-M_i^2+iM_i\Gamma_i} \sqrt{\frac{\Phi(\sqrt{s})}{\Phi(M_i)}},
\end{aligned}
\end{equation}
\begin{equation} \label{eq:phsp_func}
\begin{aligned}
        \Phi(\sqrt{s}) = \frac{q^3}{s}.
\end{aligned}
\end{equation}

\noindent In Eq.~(\ref{eq:bw_func}), $\mathcal{B}_i$,
$\Gamma_i^{\ee}$, $\Gamma_i$ and $M_i$ denote the resonance decay
branching fraction to the $\eta\jpsi$ final state, the partial width
of its decay to $\ee$, the full width, and the mass of the $i$-th
resonance. In Eq.~(\ref{eq:phsp_func}), $q$ is the daughter momentum
in the rest frame of its parent. The non-resonant part is parameterized
following the method of BaBar~\cite{PRD86051102} as 

\begin{equation}
\label{eq:conti_func} \begin{aligned}
        \sqrt{\sigma_{NY}(\sqrt{s})} = \sqrt{\Phi(\sqrt{s}) e^{-p_0u}p_1},
\end{aligned}
\end{equation}

\noindent where $p_0$ and $p_1$ are free parameters, and $u=\sqrt{s}
- (M_{\eta}+M_{\jpsi})$.

\begin{figure}[!htp]
    \begin{center}
    \begin{overpic}[width=0.5\textwidth]{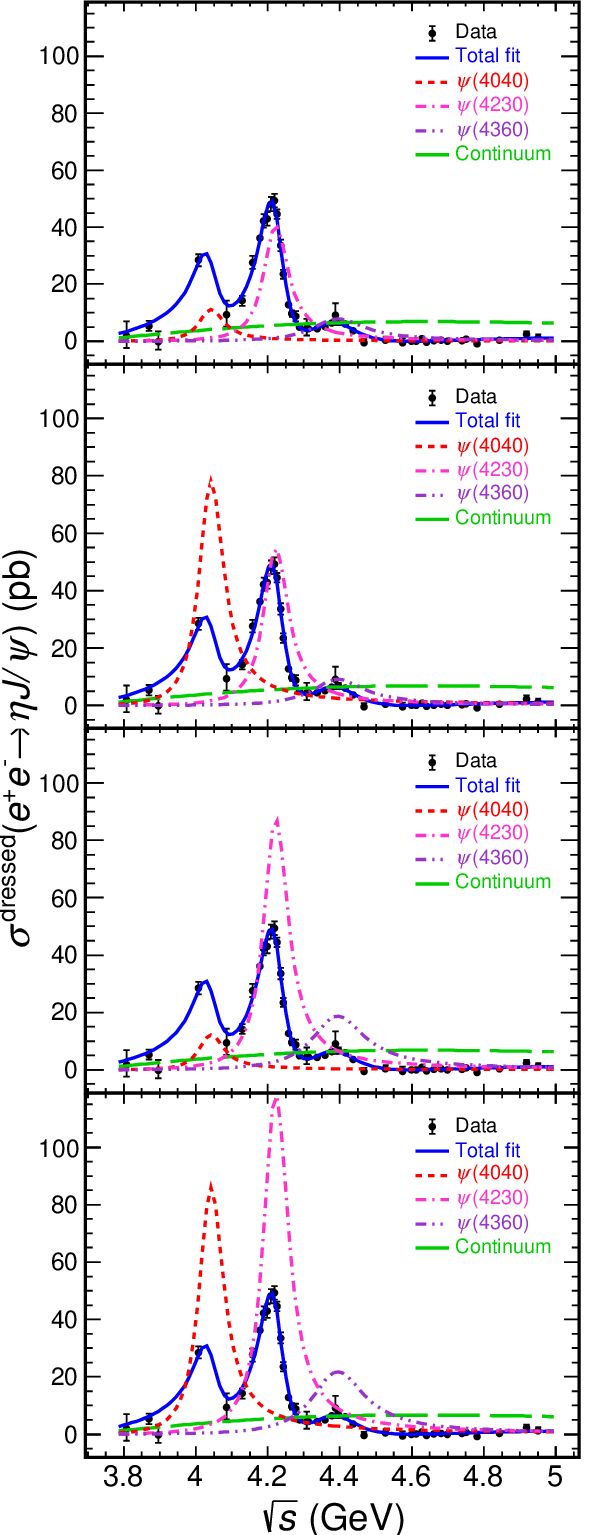}
    \put(9,96){Solution I}
    \put(9,72.3){Solution II}
    \put(9,48.6){Solution III}
    \put(9,24.8){Solution IV}
    \end{overpic}
   \end{center}
   \caption{The fits to the dressed cross sections of
   $\ee\to\eta\jpsi$ corresponding to the four solutions in
   Table~\ref{4sol_tab}. The black dots with error bars are the
   measured dressed cross sections, the blue solid curves represent
   the best fit results of the following interfering amplitudes:
   $\psi(4040)$ (dashed red), $\psi(4230)$ (short-dashed pink),
   $\psi(4360)$ (short-dashed purple), and the non-resonant component
   (long-dashed green).}

        \label{fig:lsfit}
\end{figure}

In this fit, the structure around $4040\mevcc$ is assumed to be
$\psi(4040)$. Because of the lack of data samples around
this energy region, the mass and width of the $\psi(4040)$
are fixed to the values given in the PDG~\cite{Workman:2022ynf}. By
scanning three relative phases, four solutions with similar fit
quality and identical masses and widths of the resonances around 4220
and 4390$\mevcc$ are found, consistent with the mathematical
analysis of multiple solutions shown in Ref.~\cite{Bai:2019jrb}. The
fit results are shown in Table~\ref{4sol_tab} and
Fig.~\ref{fig:lsfit}.  To estimate the significance of the three
structures and the non-resonant part, the fits are repeated removing one
of these four terms at a time.  The statistical significances of both
the non-resonant part and $\psi(4040)$ are 8.0$\sigma$, and those of the
$\psi(4230)$ and $\psi(4360)$ are more than 10.0$\sigma$.

Alternative fits are carried out by replacing the second
resonance with $\psi(4160)$ parameters and the third resonance with
$\psi(4360)$ or $\psi(4415)$ parameters from the
PDG~\cite{Workman:2022ynf}.  However, their fit qualities are
significantly worse than the nominal results and cannot describe
the data well.  To test for the existence of extra
resonances, fits are performed by adding the $\psi(4160)$ or $\psi(4415)$ 
component with the fixed parameters from the PDG~\cite{Workman:2022ynf}.
The significances of $\psi(4160)$ and $\psi(4415)$ are 3.2$\sigma$ and
1.1$\sigma$, respectively.  A fit is also performed with an
additional Breit-Wigner function with free parameters, whose
significance is 3.3$\sigma$.  In this case, the mass and width of this
extra resonance are $(4151\pm20)\mevcc$ and
$(110\pm36)\mev$, and the parameters of $\psi(4230)$ and $\psi(4360)$
turn out to be $(4226.5\pm3.3)\mevcc$ and $(56.8\pm7.4)\mev$, and
$(4412.0\pm6.9)\mevcc$ and $(82\pm20)\mev$, respectively.

	 \begin{table*}

	 \centering \caption{Results of the fits to the
	 $\ee\to\eta\jpsi$ cross sections. $M_i$, $\Gamma_{i}$,
	 $\Gamma_i^{\ee}$, $\mathcal{B}_i$ and $\phi_i$ represent the
	 parameters shown in Eq.~(\ref{eq:ls_func}),
	 Eq.~(\ref{eq:bw_func}) and Eq.~(\ref{eq:conti_func}). The
	 label $i$ = 1, 2 and 3 symbolizes $\psi(4040)$, $\psi(4230)$,
	 and $\psi(4360)$, respectively. The uncertainties are
	 statistical only.}\label{4sol_tab} 
   \begin{tabular*}{\hsize}{@{}@{\extracolsep{\fill}}lcccc@{}}
	  \hline
	  \hline
     Parameter  	& Solution I           &Solution II            &Solution III         &Solution IV\\
	\hline
	$M_{1}\  (\mevcc) $                        &\multicolumn{4}{c}{4039 (fixed)}                 \\
	$\Gamma_{1}\  (\mev)$                      &\multicolumn{4}{c}{80 (fixed)}                   \\
	$\Gamma^{\ee}_{1}\cdot\mathcal{B}_{1}\  (\eV)$    &$1.0 \pm   0.2$        &$7.1 \pm    0.6$    &$1.1\pm 0.2$    &$7.8 \pm 0.6$\\   
	$M_{2}\  (\mevcc)$		                   &\multicolumn{4}{c}{$4219.7\pm2.5$}         \\
	$\Gamma_{2}\  (\mev)$                      &\multicolumn{4}{c}{$80.7\pm4.4$}           \\
	$\Gamma^{\ee}_{2}\cdot\mathcal{B}_{2}\  (\eV)$    &$4.0  \pm   0.5$        &$5.5 \pm   0.7$    &$8.7 \pm 1.0$    &$11.9\pm1.1$\\
	$M_{3}\  (\mevcc)$	                      &\multicolumn{4}{c}{$4386.4\pm12.6$}        \\ 
	$\Gamma_{3}\  (\mev)$	                   &\multicolumn{4}{c}{$176.9\pm32.1$}         \\
	$\Gamma^{\ee}_{3}\cdot\mathcal{B}_{3}\  (\eV)$    &$1.8 \pm    0.6$        &$2.1 \pm   0.7$    &$4.3\pm 1.3$    &$5.0\pm 1.5$\\
	$\phi_{1} $\ (rad)		                      &$3.1 \pm   0.6$        &$-1.8 \pm   0.1$    &$3.3\pm 0.4$    &$-1.6\pm 0.1$\\   
	$\phi_{2} $\ (rad)	                         &$-2.8 \pm   0.1$        &$2.9 \pm   0.2$    &$-2.0\pm 0.1$    &$-2.6\pm 0.2$\\ 
	$\phi_{3} $\ (rad) 			                   &$-2.9 \pm   0.1$        &$3.0 \pm   0.1$    &$2.8\pm 0.1$    &$2.4\pm 0.7$\\
   $p_{0} \ (\mev^{-1})$                         &$1.5\pm    0.4$        &$1.5 \pm   0.4$    &$1.5\pm 0.4$       &$1.6\pm 0.4$   \\ 
	$p_{1} \ (\gev^{-3})$                         &$390.0\pm155.3$        &$389.3\pm155.6$    &$389.5\pm155.1$    &$389.5\pm154.5$\\ 
	\hline                        
	  \hline 
	\end{tabular*}
	\end{table*}

\section{SYSTEMATIC UNCERTAINTY}  
   \subsection{Systematic uncertainties for cross section measurement}
   The following sources of systematic uncertainties are considered in
the cross section measurement listed in Appendix~\ref{app:sys}. The
   uncertainty of the integrated luminosity is estimated to be 1.0\%
   using large-angle Bhabha scattering
   events~\cite{luminosity0,luminosity1,luminosity2ecms}. The
   uncertainty of the charged track reconstruction efficiency is
   estimated to be 1.0\% for each lepton~\cite{PRL110252001}. The
   charged pion is only reconstructed in Mode II. The
   uncertainty from the pion pair reconstruction efficiency for Mode
   II is 2.0\%,~\cite{BESIII:2018ldc}.  The
   uncertainty of the reconstruction efficiency per photon is
   estimated to be 1.0\%~\cite{PRD81052005}. The uncertainties of the
   branching fractions of the intermediate decays are taken from the
   PDG~\cite{Workman:2022ynf}. The uncertainty of the radiative
   correction includes two parts. The first part stems from the
   precision of the ISR calculation in the generator {\sc kkmc}. The
   other part stems from $(1+\delta^{\rm ISR})$ and $\epsilon$ in
   Eq.~(\ref{eq:born_xsec}), and depends on the input line shape of
   the cross section. Therefore, in order to estimate the uncertainty
   related to ISR correction, $(1+\delta^{\rm ISR})\cdot\epsilon$
   is evaluated 500 times by varying the input cross section line
   shape parameters with the uncertainties and the covariance matrix
   obtained from the nominal result. The standard deviation of the
   $(1+\delta^{\rm ISR})\cdot\epsilon$ distribution is
   considered as the systematic uncertainty.  The uncertainty
   associated with the $\jpsi$ mass requirement is estimated by
   smearing the $M(\LL)$ distribution of MC samples according to the
   resolution difference between data and MC simulation, and the
   resulting uncertainties in signal efficiencies are obtained.
   For the uncertainty from the kinematic fit, we correct the helix
   parameters of the charged tracks in the MC to match the pull distributions
   in the data~\cite{PhysRevD.87.012002} and reevaluate the selection
   efficiencies. The resulting changes of cross sections are considered
   as the systematic uncertainties. The systematic uncertainty of the
   photon-energy criteria in \mbox{Mode I} is considered by the
   ``Barlow-test", following the procedure described in
   Refs.~\cite{Barlow:2002yb,BESIII:2021ypr}. The uncertainties
   related to the fit procedure are estimated by changing the fit
   range, replacing the first-order polynomial function by a
   second-order polynomial function for the background description, and
   varying the width of the convolved Gaussian function for the signal shape by one
   standard deviation. The uncertainties from the
   other selections, trigger simulation, event start time
   determination, and final-state-radiation simulation and other
   sources, are conservatively taken as 1.0\%.  Assuming all sources
   of systematic uncertainties to be independent, the total
   uncertainties in the $\ee\to\eta\jpsi$ cross sections are assigned
   as the quadratic sum of the individual items, which are
   $3.8\%\sim27.9\%$ and shown in Appendix~\ref{app:Ysys}.

   \subsection{Systematic uncertainties for resonance parameters}

The systematic uncertainties for the resonance parameters in the cross
section fit are as follows. The systematic uncertainty
associated with the collision energy is conservatively estimated to be
$0.8\mev$~\cite{BESIII:2015zbz,CPC45103001}.  It is common for all
data samples and causes a global uncertainty of the mass measurement
of $Y$ states. The uncertainty due to the energy spread is estimated
by convolving the fit formula with a Gaussian function with a width of
$1.6\mev$, which is the energy spread determined by the beam energy
measurement system~\cite{Abakumova:2011rp}. The uncertainties
associated with the cross section measurement are estimated by
incorporating the correlated and uncorrelated systematic uncertainties
of the measured cross sections in the fit as shown in
Appendix~\ref{app:Ysys}.  The uncertainties from the $\psi(4040)$
resonance parameters are studied by varying the parameters within
their uncertainties from the PDG~\cite{Workman:2022ynf}.  To estimate
the uncertainty related to the parameterization of the non-resonant part,
we replace its amplitude in Eq.~(\ref{eq:conti_func}) with 
$\frac{C_0}{s^n}\sqrt{\Phi(\sqrt s)}$, where $C_0$ is a free parameter
and $\Phi(\sqrt s)$ is defined in Eq.~(\ref{eq:phsp_func}). To
estimate the uncertainty from the parameterization of the Breit-Wigner
function, $\Gamma_i$ is set to the mass dependent width $\Gamma_i =
\Gamma_i^0\cdot\frac{\Phi(\sqrt s)}{\Phi(M_i)}$, where $\Gamma_i^0$ is
the nominal width of the resonance.  We perform the fit to the cross
section line shape with the above scenarios individually, and the
resultant differences are taken as the systematic uncertainties,
listed in Table~\ref{sys_unc_resonance}. The total systematic
uncertainty is obtained by summing all sources of systematic
uncertainties in quadrature, under the assumption that they are
uncorrelated.

    \begin{table*}
    \centering
    \caption{Systematic uncertainties of resonance parameters,
      including the c.m. energy ($\sqrt s$), the energy spread ($\sqrt
      s$ spread), the $\psi(4040)$ parameters ($\psi(4040)$), the
      systematic uncertainty in the cross section measurement (Cross
      section), the parameterization of non-resonant amplitude (Fit
      model), and the parameterization of Breit-Wigner function
      ($\Gamma_{\mathrm{tot}}$). The symbol ``$\cdots$" represents
      that the uncertainty is neglected. The label $i$ = 1, 2 and 3
      symbolizes $\psi(4040)$, $\psi(4230)$, and $\psi(4360)$,
      respectively.}\label{sys_unc_resonance}
    \setlength\tabcolsep{10pt}
   \begin{tabular*}{\hsize}{@{}@{\extracolsep{\fill}}lcccccccc@{}}
     \hline
     \hline
     Source                   &Solution  &$\sqrt s$ &$\sqrt s$ spread &$\psi(4040)$ &Cross section &Fit model &$\Gamma_{\mathrm{tot}}$  &Total\\
   \hline
   $M_{2}\  (\mevcc)   $       &-  &0.8                &0.7          &0.3         &0.7          &0.2               &4.3                 &4.5 \\
   $\Gamma_{2}\  (\mev)$       &-  &$\cdots$                  &1.1          &0.3         &0.9          &0.2               &0.3                 &1.4 \\
	$M_{3}\  (\mevcc)   $       &-  &0.8                &0.4          &1.1         &0.8          &0.1               &16.8                &16.9\\ 
	$\Gamma_{3}\  (\mev)$       &-  &$\cdots$                  &9.9          &0.8         &6.7          &4.7               &2.0                 &13.0\\
	  %\hline                                                                                                                 
\multirow{3}{*}{$\Gamma^{\ee}_{1}\cdot\mathcal{B}_{1}\  (\eV)$} 
                               &I    &$\cdots$              &0.05         &0.09        &0.04          &0.01              &0.05            &0.12\\   %1229
                               &II   &$\cdots$              &0.03         &0.87        &0.04          &0.01              &0.31            &0.93\\   %1233
                               &III  &$\cdots$              &0.05         &0.11        &0.05          &0.01              &0.06            &0.15\\   %1826
                               &IV   &$\cdots$              &0.04         &1.06        &0.03          &0.01              &0.38            &1.13\\   %1826
	  %\hline                                                                                                             
\multirow{3}{*}{$\Gamma^{\ee}_{2}\cdot\mathcal{B}_{2}\  (\eV)$} 
                               &I    &$\cdots$              &0.02         &0.03        &0.10          &0.02              &0.01            &0.11\\
                               &II   &$\cdots$              &0.12         &0.26        &0.06          &0.00              &0.10            &0.31\\
                               &III  &$\cdots$              &0.18         &0.05        &0.34          &0.12              &0.03            &0.41\\
                               &IV   &$\cdots$              &0.05         &0.40        &0.30          &0.10              &0.23            &0.57\\
	  %\hline                                                                                                               
\multirow{3}{*}{$\Gamma^{\ee}_{3}\cdot\mathcal{B}_{3}\  (\eV)$} 
                               &I    &$\cdots$              &0.18         &0.00        &0.16          &0.09              &0.01            &0.26\\
                               &II   &$\cdots$              &0.22         &0.04        &0.18          &0.10              &0.02            &0.30\\
                               &III  &$\cdots$              &0.30         &0.05        &0.30          &0.16              &0.01            &0.45\\
                               &IV   &$\cdots$              &0.36         &0.14        &0.34          &0.17              &0.03            &0.54\\
     \hline                                                
     \hline                                                
   \end{tabular*}
   \end{table*} 

\section{Summary and discussion}

In summary, we measure the cross sections of $\ee\to\eta\jpsi$ at
c.m. energies between $3.808$ and $4.951\gev$ using data samples with
an integrated luminosity of 22.42\,$\ifb$ collected by the BESIII
detector operating at the BEPCII collider. The measured Born cross
sections are consistent with the previous BESIII
measurements~\cite{PRD86071101,PRD91112005,PRD102031101}. However,
additional cross sections are measured on both sides of the
$\psi(4230)$ peak, and from $4.612$ to $4.951\gev$, allowing the line
shape to be studied more precisely than before.

The dressed cross sections are
fitted with the three resonance amplitudes and a non-resonant
amplitude. Assuming the lowest lying structure as $\psi(4040)$, the
$\psi(4230)$ and $\psi(4360)$ structures are clearly observed with
statistical significance much greater than $10.0\sigma$. The masses
and widths of these two states are determined as $M =
(4219.7\pm2.5\pm4.5)\mevcc$, $\Gamma = (80.7 \pm 4.4 \pm1.4)\mev$ for
$\psi(4230)$, and $M = (4386\pm13\pm17)\mevcc$, $\Gamma =
(177\pm32\pm13)\mev$ for $\psi(4360)$, respectively.  A comparison of
the parameters of $\psi(4230)$ and $\psi(4360)$ obtained in this
analysis and the previous \mbox{BESIII} ones is shown in
Fig.~\ref{fig:Y_par_cmp}. The parameters of $\psi(4360)$ are
consistent within uncertainties. However, the width of $\psi(4230)$
obtained in this analysis is larger than those obtained in other
processes~\cite{BESIII:2020oph,BESIII:2022qal,PRD104052012,PRD99091103}.

Based on the four solutions including the statistical and systematic
uncertainties and combining with the electronic
partial widths, which are $0.63\sim0.66\kev$ for $\psi(4230)$ and
$0.523\kev$ for $\psi(4360)$ in Refs.~\cite{PRD91094023,PRD95114031},
the branching fraction $\mathcal{B}(\psi(4230)\to\eta\jpsi)$ is
estimated to be in the range of $(6.06\pm0.76\pm0.17)\times10^{-3}$
to $(18.89\pm1.75\pm0.90)\times10^{-3}$, and the partial decay width
$\Gamma(\psi(4360)\to\eta\jpsi)$ is estimated to be in the range of
$(0.61\pm0.23\pm0.10)\mev$ to $(1.70\pm0.59\pm0.22)\mev$. But neither
of them can cover the predictions of
Refs.~\cite{PRD91094023,PRD95114031} based on a conventional charmonium
state model.  Comparing with $\Gamma^{\ee}_{\psi(4360)}\cdot
\mathcal{B}(\psi(4360)\to\pppm h_c)$ from Ref.~\cite{PRL118092002}, we
obtain the ratio
$\frac{\Gamma(\psi(4360)\to\etajpsi)}{\Gamma(\psi(4360)\to\pppm h_c)}
= 0.16^{+0.08}_{-0.07}\pm0.03\sim 0.43^{+0.23}_{-0.21}\pm0.08$, which
is beyond the expected range under the $D^{*}\bar{D_1}$ +
H.c.$\footnote{Hermitian conjugate}$ molecular scenario in
Ref.~\cite{PhysRevD.96.054017}.  Further theoretical and experimental
studies are still needed to interpret the nature and the structures of
these states.

\begin{figure}
   \includegraphics[width=3.25in]{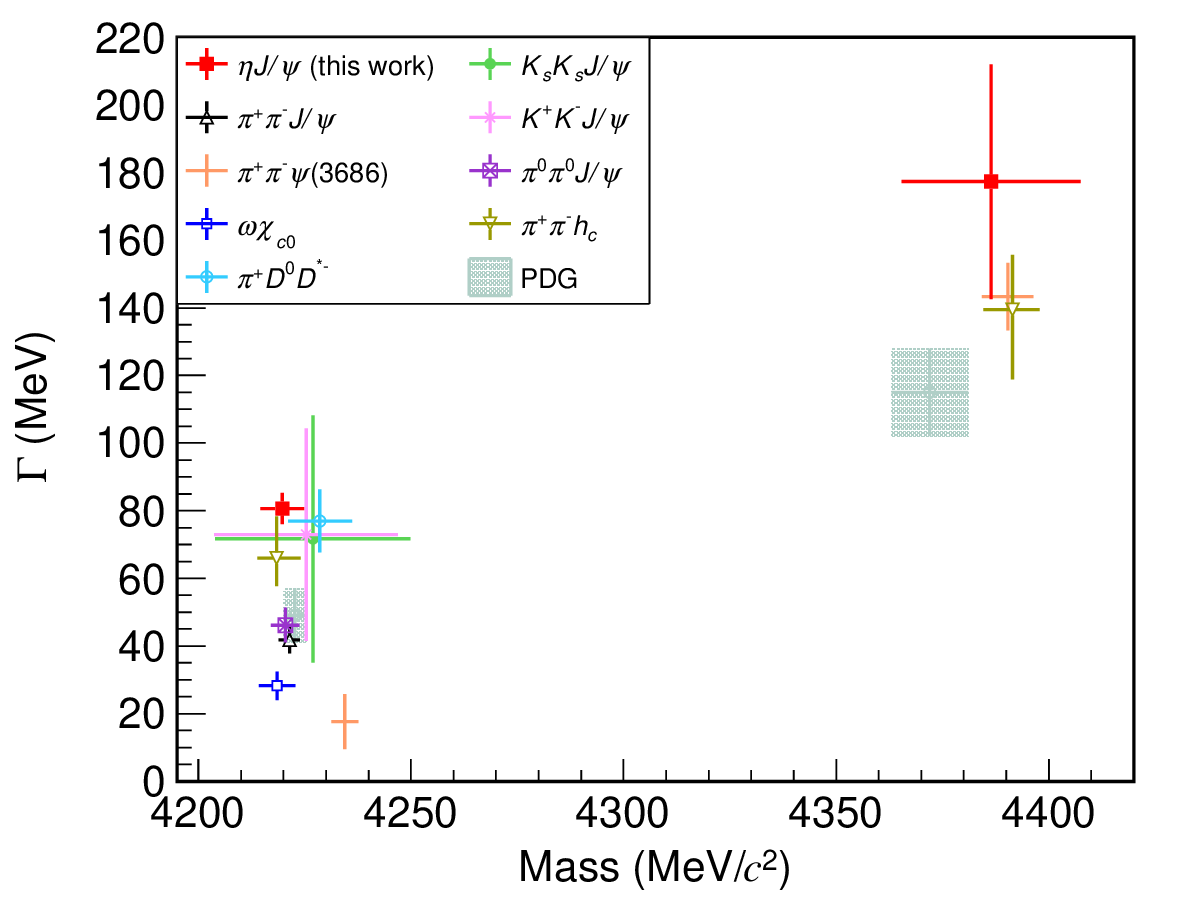}
   \caption{Comparison of masses versus widths of $\psi(4230)$ and $\psi(4360)$ from the previous BESIII measurements~\cite{PhysRevD.107.092005,BESIII:2022qal,BESIII:2022joj,PRD104052012,BESIII:2020oph,PRD99091103,PRL118092002,BESIII:2018iea} and the average values in the PDG~\cite{Workman:2022ynf}. The results in the bottom left are for $\psi(4230)$, and the ones in the top right are for $\psi(4360)$. \label{fig:Y_par_cmp}}
\end{figure} 
%\clearpage

\begin{acknowledgments}
%\section{ACKNOWLEDGMENTS}
The BESIII collaboration thanks the staff of BEPCII and the IHEP computing center and the supercomputing center of the University of Science and Technology of China (USTC) for their strong support. 
This work is supported in part by National Key R\&D Program of China under Contracts Nos. 2020YFA0406400, 2020YFA0406300; National Natural Science Foundation of China (NSFC) under Contracts Nos. 11335008, 11625523, 12035013, 11705192, 11950410506, 12061131003, 12105276, 12122509, 11635010, 11735014, 11835012, 11935015, 11935016, 11935018, 11961141012, 12022510, 12025502, 12035009, 12192260, 12192261, 12192262, 12192263, 12192264, 12192265, 12221005, 12225509, 12235017, 12005311; the Chinese Academy of Sciences (CAS) Large-Scale Scientific Facility Program; the CAS Center for Excellence in Particle Physics (CCEPP); Joint Large-Scale Scientific Facility Funds of the NSFC and CAS under Contract No. U1732263, U1832103, U2032111, U1832207; CAS Key Research Program of Frontier Sciences under Contracts Nos. QYZDJ-SSW-SLH003, QYZDJ-SSW-SLH040; 100 Talents Program of CAS; The Institute of Nuclear and Particle Physics (INPAC) and Shanghai Key Laboratory for Particle Physics and Cosmology; European Union's Horizon 2020 research and innovation programme under Marie Sklodowska-Curie grant agreement under Contract No. 894790; German Research Foundation DFG under Contracts Nos. 455635585, Collaborative Research Center CRC 1044, FOR5327, GRK 2149; Istituto Nazionale di Fisica Nucleare, Italy; Ministry of Development of Turkey under Contract No. DPT2006K-120470; National Research Foundation of Korea under Contract No. NRF-2022R1A2C1092335; National Science and Technology fund of Mongolia; National Science Research and Innovation Fund (NSRF) via the Program Management Unit for Human Resources \& Institutional Development, Research and Innovation of Thailand under Contract No. B16F640076; Polish National Science Centre under Contract No. 2019/35/O/ST2/02907; The Swedish Research Council; U. S. Department of Energy under Contract No. DE-FG02-05ER41374.
\end{acknowledgments}

\appendix
\section{SUMMARY OF THE BORN CROSS SECTIONS} \label{app:cs}
%Table~\ref{cs} summarizes physical variables used in the Born cross section (or the upper limit at the 90\% confidence level) calculation.
Table~\ref{cs} shows the values relating to the details of the calculation of the Born section. $\sigma^{UL}$ denotes an estimate of the upper limit of the cross section at the 90\% confidence level.
\begin{table*}
%        \scriptsize
   \centering
   \caption{The c.m. energy, the corresponding luminosity $\mathcal{L}$, the detection efficiency $\epsilon^i$, the ISR correction factor $1+\delta^{\rm ISR}$, vacuum polarization factor $\frac{1}{|1-\Pi|^2}$, and the obtained Born cross section at each energy point, where the first uncertainties for cross sections are statistical and the second systematic.}\label{cs}
      \begin{threeparttable}
   \begin{tabular*}{\hsize}{@{}@{\extracolsep{\fill}}ccccccccccc@{}}
      \hline
      \hline
 $\sqrt s~(\gev)$
& $\mathcal{L}~(\ipb)$
& $\epsilon~(\%)$ \tnote{1}
& $\epsilon~(\%)$ \tnote{2}
& $\epsilon~(\%)$ \tnote{3}
& $\epsilon~(\%)$ \tnote{4}
& $1+\delta^{\rm ISR}$
& $\frac{1}{|1-\Pi|^2}$
& $\sigma~({\rm pb})$
& $\sigma^{UL}~({\rm pb})$ \\
      \hline
3.8077&  50.54  & 25.13&  17.27&   9.56&   6.71&  1.598    &  1.056&  $1.51  ^{+4.18 }_{-3.41} \pm0.06$& 8.71\\
%3.8077&  50.54  & 43.16&  29.97&  18.54&  12.86&  0.787    &  1.056&  $1.74  ^{+4.82 }_{-3.94} \pm0.07$& 10.04\\
3.8694&  219.2 &  41.67&  29.24&  17.66&  12.40&  0.801&      1.051&  $5.07 \pm1.69 \pm0.20$&         -       \\
3.8962&  52.61  & 41.66&  29.37&  17.49&  12.73&  0.802    &  1.049&  $-0.17 ^{+3.38 }_{-2.59} \pm0.01$& 6.53 \\
4.0076&  482.0 &  42.21&  29.96&  17.96&  13.08&  0.791&      1.044&  $27.26\pm2.03 \pm1.06$&         -       \\
4.0855&  52.86  & 35.51&  25.12&  15.02&  10.64&  1.007    &  1.051&  $8.87  ^{+4.73 }_{-3.97} \pm0.36$& 15.97\\
4.1285&  401.5 &  36.87&  26.20&  15.21&  11.10&  0.896&      1.052&  $13.47\pm1.61 \pm0.54$&         -       \\
4.1574&  408.7 &  39.35&  27.80&  16.20&  11.73&  0.831&      1.053&  $26.24\pm2.13 \pm1.02$&         -       \\
4.1784&  3194.5&  40.17&  28.23&  16.63&  11.88&  0.804&      1.054&  $34.29\pm0.87 \pm1.30$&         -       \\
4.1888&  570.03&  40.64&  28.59&  16.82&  12.20&  0.796&      1.056&  $39.97\pm2.19 \pm1.52$&         -       \\
4.1989&  526.0 &  41.08&  29.06&  17.13&  12.24&  0.795&      1.056&  $40.66\pm2.21 \pm1.55$&         -       \\
4.2091&  572.05&  40.79&  28.67&  16.74&  11.82&  0.805&      1.057&  $45.54\pm2.28 \pm1.73$&         -       \\
4.2186&  569.2 &  40.38&  28.60&  16.52&  12.08&  0.829&      1.056&  $46.67\pm2.22 \pm1.82$&         -       \\
4.2263&  1100.9&  40.43&  28.44&  16.55&  12.12&  0.863&      1.056&  $42.15\pm1.55 \pm1.64$&         -       \\
4.2357&  530.3 &  39.07&  27.41&  15.83&  11.42&  0.927&      1.056&  $31.82\pm1.87 \pm1.24$&         -       \\
4.2436&  593.98&  37.54&  26.26&  15.06&  10.78&  1.001&      1.056&  $22.24\pm1.51 \pm0.89$&         -       \\
4.2580&  828.4 &  33.28&  23.16&  13.36&  9.63&   1.198&      1.054&  $12.10\pm0.97 \pm0.51$&         -       \\
4.2668&  531.1 &  30.00&  20.93&  11.80&  8.37&   1.358&      1.053&  $9.19 \pm1.05 \pm0.39$&         -       \\
4.2777&  175.7 &  25.80&  18.11&  9.92&   7.11&   1.585&      1.053&  $8.33 \pm1.71 \pm0.38$&         -       \\
4.2879&  502.4 &  22.89&  15.90&  8.56&   6.21&   1.774&      1.053&  $4.68 \pm0.86 \pm0.21$&         -       \\
4.3079&  45.08  & 20.15&  14.04&  7.71&   5.59&   1.835    &  1.052&  $4.25  ^{+3.22 }_{-2.41} \pm0.19$& 9.59 \\
4.3121&  501.2 &  20.17&  14.33&  7.65&   5.54&   1.787&      1.052&  $4.23 \pm0.84 \pm0.18$&         -       \\
4.3374&  505.0  & 23.19&  16.59&  9.02&   6.49&   1.430    &  1.051&  $4.19  ^{+1.00 }_{-0.92} \pm0.17$& 5.61 \\
4.3583&  543.9 &  26.53&  18.63&  10.41&  7.47&   1.241&      1.051&  $4.84 \pm0.90 \pm0.19$&         -       \\
4.3774&  522.7 &  28.58&  19.75&  10.83&  7.92&   1.171&      1.051&  $6.09 \pm0.99 \pm0.24$&         -       \\
4.3874&  55.57  & 28.72&  20.24&  11.07&  8.06&   1.165    &  1.051&  $8.70  ^{+4.06 }_{-3.29} \pm0.34$& 14.88\\
4.3965&  507.8 &  28.72&  20.40&  10.81&  7.83&   1.174&      1.051&  $6.45 \pm1.02 \pm0.26$&         -       \\
4.4156&  1090.7&  27.71&  19.42&  10.42&  7.53&   1.240&      1.052&  $5.74 \pm0.65 \pm0.23$&         -       \\
4.4362&  569.9 &  25.48&  18.08&  9.35&   6.74&   1.391&      1.054&  $3.42 \pm0.72 \pm0.14$&         -       \\
4.4671&  111.09 & 19.78&  13.98&  7.10&   5.12&   1.836    &  1.055&  $-0.50 ^{+1.26 }_{-0.85} \pm0.02$& 2.41 \\
4.5271&  112.12 & 7.44&   5.18&   2.48&   1.80&   5.347    &  1.054&  $0.40  ^{+1.17 }_{-0.86} \pm0.02$& 2.59 \\
4.5745&  48.93  & 1.23&   0.85&   0.37&   0.27&   43.404   &  1.055&  $-0.52 ^{+1.46 }_{-0.89} \pm0.08$& 3.28 \\
4.5995&  586.9  & 0.18&   0.12&   0.04&   0.03&   19424.100&  1.055&  $-0.007^{+0.004}_{-0.003}\pm0.00$& 0.01 \\
4.6119&  103.83 & 0.21&   0.13&   0.05&   0.03&   276.006  &  1.055&  $0.01  ^{+0.77 }_{-0.52} \pm0.00$& 1.78 \\
4.6280&  521.52 & 0.64&   0.45&   0.19&   0.14&   48.808   &  1.054&  $0.79  ^{+0.68 }_{-0.58} \pm0.04$& 1.82 \\
4.6409&  552.41 & 1.31&   0.92&   0.40&   0.29&   23.729   &  1.054&  $-0.37 ^{+0.53 }_{-0.43} \pm0.02$& 0.84 \\
4.6612&  529.63 & 2.68&   1.93&   0.82&   0.61&   11.549   &  1.054&  $0.12  ^{+0.55 }_{-0.46} \pm0.00$& 1.09 \\
4.6819&  1669.31& 4.49&   3.15&   1.35&   0.97&   7.118    &  1.054&  $0.34  ^{+0.33 }_{-0.30} \pm0.01$& 0.83 \\
4.6988&  536.45 & 6.08&   4.20&   1.77&   1.30&   5.345    &  1.055&  $-0.01 ^{+0.56 }_{-0.46} \pm0.00$& 1.03 \\
4.7397&  164.27 & 9.85&   6.92&   2.84&   2.02&   3.335    &  1.055&  $0.21  ^{+0.89 }_{-0.63} \pm0.01$& 1.94 \\
4.7501&  367.21 & 10.85&  7.63&   3.04&   2.23&   3.046    &  1.055&  $0.62  ^{+0.76 }_{-0.64} \pm0.02$& 1.83 \\
4.7805&  512.78 & 13.34&  9.45&   3.67&   2.68&   2.488    &  1.055&  $-0.85 ^{+0.43 }_{-0.31} \pm0.03$& 0.60 \\
4.8431&  527.29 & 17.72&  12.67&  4.57&   3.32&   1.889    &  1.056&  $0.31  ^{+0.52 }_{-0.43} \pm0.01$& 1.16 \\
4.9180&  208.11 & 21.37&  15.16&  4.98&   3.68&   1.575    &  1.056&  $2.18  ^{+1.09 }_{-0.89} \pm0.08$& 3.86 \\
4.9509&  160.37 & 22.31&  15.89&  5.04&   3.63&   1.496    &  1.056&  $1.08  ^{+1.21 }_{-0.92 }\pm0.04$& 3.11 \\
      \hline
      \hline
   \end{tabular*}
        \begin{tablenotes}
        \footnotesize
        \item[1] $\jpsi \to \mumu,\eta \to \gamma\gamma$
        \item[2] $\jpsi \to \ee,\eta \to \gamma\gamma$
        \item[3] $\jpsi \to \mumu,\eta \to\piz\pppm$
        \item[4] $\jpsi \to \ee,\eta \to\piz\pppm$
       \end{tablenotes}
    \end{threeparttable}
\end{table*}  

%\clearpage
\section{SYSTEMATIC UNCERTAINTIES ON THE CROSS SECTIONS} \label{app:sys}
All systematic uncertainties at individual c.m. energies are summarized in Table~\ref{sys}.
      The sources with the symbol ``*" are the correlated systematic uncertainties for different data samples.
      Due to the limited statistics of most data samples, the items with the symbol ``$\dag$", are estimated with the data sample with the highest statistics ($\sqrt s = 4.226\gev$).
      The total systematic uncertainties are obtained with the quadrature sum of individual uncertainties by assuming all of them are independent.
 \begin{table*}
 %\begin{table*}[H]
 \centering
 \caption{Relative systematic uncertainties ($\%$) in the Born cross section measurement. The sources with ``*" are the common systematic uncertainties for different c.m. energies. The items with the symbol ``$\dag$" are estimated with the data sample with the highest statistics ($\sqrt s = 4.226\gev$). The systematic uncertainties include Luminosity~($\mathcal{L}_{\rm int}$), Branching fraction ($\mathcal B$), ISR correction, $\gamma$ detection, Tracking~($\mumu/\EE$)~(Track1), Tracking~($\pppm$)~(Track2), Lepton pair mass window~$M(\LL)$, Kinematic fit, Photon-energy criteria~($E_{\gamma}$), Fit~(including background shape, fit range and signal shape), and Others.}\label{sys}
%\scriptsize
   \begin{tabular*}{\hsize}{@{}@{\extracolsep{\fill}}ccccccccccccc@{}}
  \hline
  \hline
  $\sqrt{s}~(\gev)$&  $\mathcal{L}_{\rm int}$*& $\mathcal B$*& ISR correction& $\gamma$ detection*$^{\dag}$& Track1*&   Track2*& $M(\LL)^{\dag}$&  Kinematic fit$^{\dag}$&    $E_{\gamma}$&  Fit$^{\dag}$&  Others*&  Total\\
  \hline
3.8077&              1.0&                   0.9&        0.5&              2.0&              2.0&        0.4&       0.1&           0.8&       1.2&         1.0&   1.0&       3.8 \\
3.8694&              1.0&                   0.9&        0.5&              2.0&              2.0&        0.9&       0.1&           0.8&       1.2&         1.0&   1.0&       3.9 \\
3.8962&              1.0&                   0.9&        0.5&              2.0&              2.0&        0.5&       0.1&           0.8&       1.2&         1.0&   1.0&       3.8 \\
4.0076&              1.0&                   0.9&        0.7&              2.0&              2.0&        0.5&       0.1&           0.8&       1.2&         1.0&   1.0&       3.9 \\
4.0855&              1.0&                   0.9&        1.2&              2.0&              2.0&        0.4&       0.1&           0.8&       1.2&         1.0&   1.0&       4.1 \\
4.1285&              1.0&                   0.9&        1.1&              2.0&              2.0&        0.5&       0.1&           0.8&       1.2&         1.0&   1.0&       4.0 \\
4.1574&              1.0&                   0.9&        0.7&              2.0&              2.0&        0.5&       0.1&           0.8&       1.2&         1.0&   1.0&       3.9 \\
4.1784&              1.0&                   0.9&        0.6&              2.0&              2.0&        0.5&       0.1&           0.8&       1.2&         1.0&   1.0&       3.8 \\
4.1888&              1.0&                   0.9&        0.6&              2.0&              2.0&        0.5&       0.1&           0.8&       1.2&         1.0&   1.0&       3.8 \\
4.1989&              1.0&                   0.9&        0.6&              2.0&              2.0&        0.5&       0.1&           0.8&       1.2&         1.0&   1.0&       3.8 \\
4.2091&              1.0&                   0.9&        0.6&              2.0&              2.0&        0.4&       0.1&           0.8&       1.2&         1.0&   1.0&       3.8 \\
4.2186&              1.0&                   0.9&        0.6&              2.0&              2.0&        0.4&       0.1&           0.8&       1.2&         1.0&   1.0&       3.9 \\
4.2263&              1.0&                   0.9&        0.7&              2.0&              2.0&        0.4&       0.1&           0.8&       1.2&         1.0&   1.0&       3.9 \\
4.2357&              1.0&                   0.9&        0.8&              2.0&              2.0&        0.4&       0.1&           0.8&       1.2&         1.0&   1.0&       3.9 \\
4.2436&              1.0&                   0.9&        0.9&              2.0&              2.0&        0.5&       0.1&           0.8&       1.2&         1.0&   1.0&       4.0 \\
4.2580&              1.0&                   0.9&        1.4&              2.0&              2.0&        0.5&       0.1&           0.8&       1.2&         1.0&   1.0&       4.2 \\
4.2668&              1.0&                   0.9&        1.6&              2.0&              2.0&        0.5&       0.1&           0.8&       1.2&         1.0&   1.0&       4.3 \\
4.2777&              1.0&                   0.9&        2.2&              2.0&              2.0&        0.5&       0.1&           0.8&       1.2&         1.0&   1.0&       4.6 \\
4.2879&              1.0&                   0.9&        2.0&              2.0&              2.0&        0.6&       0.1&           0.8&       1.2&         1.0&   1.0&       4.5 \\
4.3079&              1.0&                   0.9&        2.0&              2.0&              2.0&        0.4&       0.1&           0.8&       1.2&         1.0&   1.0&       4.5 \\
4.3121&              1.0&                   0.9&        1.7&              2.0&              2.0&        0.6&       0.1&           0.8&       1.2&         1.0&   1.0&       4.3 \\
4.3374&              1.0&                   0.9&        1.2&              2.0&              2.0&        0.4&       0.1&           0.8&       1.2&         1.0&   1.0&       4.1 \\
4.3583&              1.0&                   0.9&        0.9&              2.0&              2.0&        0.6&       0.1&           0.8&       1.2&         1.0&   1.0&       4.0 \\
4.3774&              1.0&                   0.9&        0.9&              2.0&              2.0&        0.5&       0.1&           0.8&       1.2&         1.0&   1.0&       4.0 \\
4.3874&              1.0&                   0.9&        0.8&              2.0&              2.0&        0.4&       0.1&           0.8&       1.2&         1.0&   1.0&       3.9 \\
4.3965&              1.0&                   0.9&        0.9&              2.0&              2.0&        0.5&       0.1&           0.8&       1.2&         1.0&   1.0&       4.0 \\
4.4156&              1.0&                   0.9&        0.9&              2.0&              2.0&        0.5&       0.1&           0.8&       1.2&         1.0&   1.0&       4.0 \\
4.4362&              1.0&                   0.9&        1.1&              2.0&              2.0&        0.6&       0.1&           0.8&       1.2&         1.0&   1.0&       4.0 \\
4.4671&              1.0&                   0.9&        1.2&              2.0&              2.0&        0.3&       0.1&           0.8&       1.2&         1.0&   1.0&       4.1 \\
4.5271&              1.0&                   0.9&        2.1&              2.0&              2.0&        0.3&       0.1&           0.8&       1.2&         1.0&   1.0&       4.5 \\
4.5745&              1.0&                   0.9&        14.9&             2.0&              2.0&        0.5&       0.1&           0.8&       1.2&         1.0&   1.0&       15.9\\
4.5995&              1.0&                   0.9&        27.2&             2.0&              2.0&        0.2&       0.1&           0.8&       1.2&         1.0&   1.0&       27.9\\
4.6119&              1.0&                   0.9&        17.9&             2.0&              2.0&        0.5&       0.1&           0.8&       1.2&         1.0&   1.0&       18.8\\
4.6280&              1.0&                   0.9&        2.3&              2.0&              2.0&        0.3&       0.1&           0.8&       1.2&         1.0&   1.0&       4.6 \\
4.6409&              1.0&                   0.9&        1.2&              2.0&              2.0&        0.3&       0.1&           0.8&       1.2&         1.0&   1.0&       4.1 \\
4.6612&              1.0&                   0.9&        1.0&              2.0&              2.0&        0.3&       0.1&           0.8&       1.2&         1.0&   1.0&       4.0 \\
4.6819&              1.0&                   0.9&        0.8&              2.0&              2.0&        0.3&       0.1&           0.8&       1.2&         1.0&   1.0&       3.9 \\
4.6988&              1.0&                   0.9&        0.7&              2.0&              2.0&        0.3&       0.1&           0.8&       1.2&         1.0&   1.0&       3.9 \\
4.7397&              1.0&                   0.9&        0.6&              2.0&              2.0&        1.0&       0.1&           0.8&       1.2&         1.0&   1.0&       4.0 \\
4.7501&              1.0&                   0.9&        0.6&              2.0&              2.0&        0.3&       0.1&           0.8&       1.2&         1.0&   1.0&       3.8 \\
4.7805&              1.0&                   0.9&        0.6&              2.0&              2.0&        0.3&       0.1&           0.8&       1.2&         1.0&   1.0&       3.8 \\
4.8431&              1.0&                   0.9&        0.6&              2.0&              2.0&        0.3&       0.1&           0.8&       1.2&         1.0&   1.0&       3.8 \\
4.9180&              1.0&                   0.9&        0.6&              2.0&              2.0&        0.2&       0.1&           0.8&       1.2&         1.0&   1.0&       3.8 \\
4.9509&              1.0&                   0.9&        0.5&              2.0&              2.0&        0.2&       0.1&           0.8&       1.2&         1.0&   1.0&       3.8 \\
  \hline
  \hline
\end{tabular*}
\end{table*}

\section{DEFINITION OF LIKELIHOOD FUNCTION CONSIDERING THE SYSTEMATIC UNCERTAINTIES OF CROSS SECTIONS} \label{app:Ysys}
                In the maximum-likelihood fit of the dressed cross sections of $\ee\to\eta\jpsi$, to consider the systematic uncertainties of resonance parameters from the cross section measurement, the systematic uncertainties of cross section measurement are divided into two parts, uncorrelated and correlated.
                Assuming all sources to be independent, the total uncorrelated and correlated relative systematic uncertainties are obtained by adding their individual values in quadrature separately.
\begin{itemize}
           \item \textbf{Uncorrelated part}\\
              The likelihood function of the $i$-th data sample, considering the uncorrelated uncertainty from cross section measurement as the nuisance parameter following the Gaussian distribution, is defined as:
\begin{equation} \label{eq:uncorrL}
\begin{aligned}
          L'_i = \int L_i(\sigma^{\rm fit}_i\cdot \epsilon) \times Gauss(\epsilon; 1, \epsilon^{\rm uncorr}_i) d\epsilon,
\end{aligned}
\end{equation}
        where $L_i$ is the likelihood function with only statistical uncertainties, $\sigma^{\rm fit}_i$ is the expected value of cross section and $\epsilon_i^{\rm uncorr}$ is the total uncorrelated systematic uncertainty in the cross section measurement of the $i$-th data sample. 
        %$G_i$ is the asymmetric Gaussian function constructed by the measured cross section and its statistical uncertainties
      \item \textbf{Correlated part}\\
              Considering the correlated systematic uncertainties, which obey a Gaussian distribution, as the nuisance parameter of the overall likelihood function in the line shape fit, the likelihood function of total data samples is defined as:
\begin{equation} \label{eq:corrL}
\begin{aligned}
              L'_{\rm tot} = \int [\prod\limits_{i=1}^{44}L'_{i}(\sigma^{\rm fit}_i\cdot \epsilon)] \times Gauss(\epsilon; 1, \epsilon^{\rm corr}) d\epsilon,
\end{aligned}
\end{equation}
              where $\epsilon^{\rm corr}$ is the total correlated relative systematic uncertainty.
              Some correlated uncertainties are different for each energy point, to be conservative, the largest value is used.
	\end{itemize}
   Finally, using $L'_{\rm tot}$ to repeat the fit, the differences of the results are considered as the systematic uncertainties from the cross section measurement.
\clearpage	
\bibliography{sub_lib.bib}
\end{document}